\documentclass[lettersize,journal]{IEEEtran}
\usepackage{amsmath,amsfonts}
\usepackage{algorithmic}
\usepackage{algorithm}
\usepackage{array}
\usepackage[caption=false,font=normalsize,labelfont=sf,textfont=sf]{subfig}
\usepackage{textcomp}
\usepackage{stfloats}
\usepackage{url}
\usepackage{verbatim}
\usepackage{graphicx}
\usepackage{cite}

\begin{document}

\title{Intelligent Traffic Monitoring with Distributed Acoustic Sensing}

\author{Dongzi Xie, Xinming Wu, Zhixiang Guo, Heting Hong, Baoshan Wang, Yingjiao Rong

\thanks{Dongzi Xie, Xinming Wu, Zhixiang Guo, Heting Hong and Baoshan Wang are with School of Earth and 
Space Sciences, University of Science and Technology of China, Hefei, China and Mengcheng National 
Geophysical Observatory, Mengcheng, China (email: mutian@mail.ustc.edu.cn; xinmwu@ustc.edu.cn; 
zxg3@mail.ustc.edu.cn; honght@mail.ustc.edu.cn; bwgeo@ustc.edu.cn). Yingjiao Rong is with the Science 
and Technology on Near-Surface Detection Laboratory, Wuxi, China (email: enjoy\_rong@163.com). 
Corresponding author: Xinming Wu, Baoshan Wang, and Yingjiao Rong.}}

\markboth{IEEE TRANSACTIONS ON INTELLIGENT TRANSPORTATION SYSTEMS}
{Shell \MakeLowercase{\textit{et al.}}: A Sample Article Using IEEEtran.cls for IEEE Journals}


\maketitle

\begin{abstract}
Distributed Acoustic Sensing (DAS) is promising for traffic monitoring, but its extensive data and 
sensitivity to vibrations, causing noise, pose computational challenges. To address this, we propose a 
two-step deep-learning workflow with high efficiency and noise immunity for DAS-based traffic monitoring, 
focusing on instance vehicle trajectory segmentation and velocity estimation. Our approach begins by 
generating a diverse synthetic DAS dataset with labeled vehicle signals, tackling the issue of missing 
training labels in this field. This dataset is used to train a Convolutional Neural Network (CNN) to 
detect linear vehicle trajectories from the noisy DAS data in the time-space domain. However, due to 
significant noise, these trajectories are often fragmented and incomplete. To enhance accuracy, we 
introduce a second step involving the Hough transform. This converts detected linear features into 
point-like energy clusters in the Hough domain. Another CNN is then employed to focus on these energies, 
identifying the most significant points. The inverse Hough transform is applied to these points to 
reconstruct complete, distinct, and noise-free linear vehicle trajectories in the time-space domain. 
The Hough transform plays a crucial role by enforcing a local linearity constraint on the trajectories, 
enhancing continuity and noise immunity, and facilitating the separation of individual trajectories and 
estimation of vehicle velocities (indicated by trajectory slopes in the Hough domain). Our method has 
shown effectiveness in real-world datasets, proving its value in real-time processing of DAS data and 
applicability in similar traffic monitoring scenarios. All related codes and data are available at 
https://github.com/TTMuTian/itm/.
\end{abstract}


\section{Introduction}
\IEEEPARstart{W}{ith} the development of today's transportation network, a large number of infrastructure such as 
highways, railways, and high-speed railways have been built, traffic management has become more 
and more important\cite{A_Comprehensive_Study_of_Optical_Fiber_Acoustic_Sensing}. There are 
currently many technical solutions for traffic monitoring, such as magnetic sensors, cameras, radar 
sensors, ultrasonic waves, vehicle-mounted GPS, etc.\cite{Sensor_Technologies_for_Intelligent_Transportation_Systems}. 
However, the above technical solutions have a series of limitations, such as high deployment and 
maintenance costs, low spatio-temporal resolution, susceptibility to extreme weather, and privacy 
issues\cite{Self-Supervised_Velocity_Field_Learning_for_High-Resolution_Traffic_Monitoring_with_Distributed_Acoustic_Sensing}, 
therefore, we need an efficient and economical method for traffic monitoring.

Distributed Acoustic Sensing (DAS) is a fiber-based sensing technology that uses coherent light 
backscattered from the sensing fiber to detect and analyze disturbances to the sensing fiber caused 
by external physical fields, which can be used to measure temperature distribution, strain or 
vibro-acoustic properties of various objects. DAS has the advantages of low average cost, high 
temporal and spatial resolution, and anti-electromagnetic interference. Therefore, it has a wide 
range of applications in scenarios such as perimeter security\cite{Rapid_Surface_Deployment_of_a_DAS_System_forEarthquake_Hazard_Assessment}, 
seismic observation\cite{Ground_Motion_Response_to_an_ML_4.3_Earthquake_Using_Co-located_Distributed_Acoustic_Sensing_and_Seismometer_Arrays, 
Preliminary_Study_on_Data_Recorded_by_a_3D_Distributed_Acoustic_Sensing_Array_in_the_Huainandeep_Underground_Laboratory}, 
and underground structure detection\cite{Utilizing_Distributed_Acoustic_Sensing_and_Ocean_Bottom_Fiber_Optic_Cables_For_Submarine_Structural_Characterization, 
Distributed_Sensing_of_Microseisms_and_Teleseisms_with_Submarine_Dark_Fibers}, \cite{Hydraulic-fracture_geometry_characterization_using_low-frequency_DAS_signal}.

Due to the above advantages of DAS, it has been widely applied in traffic monitoring across various 
modes of transportation, including trains\cite{Real-Time_Train_Tracking_from_Distributed_Acoustic_Sensing_Data, 
Fiber_Optic_Sensing_Technology_and_Vision_Sensing_Technology_for_Structural_Health_Monitoring}, 
cars\cite{Vehicle_Detection_and_Classification_Using_Distributed_Fiber_Optic_Acoustic_Sensing, 
Long-Range_Traffic_Monitoring_Based_on_Pulse-Compression_Distributed_Acoustic_Sensing_and_Advanced_Vehicle_Tracking_and_Classification_Algorithm}, 
boats\cite{Preliminary_assessment_of_ship_detection_and_trajectory_evaluation_using_distributed_acoustic_sensing_on_an_optical_fiber_telecom_cable}, 
and even pedestrians\cite{Footstep_detection_in_urban_seismic_data_with_a_convolutional_neural_network}. 
In the research on vehicle monitoring, some studies detect vehicles by analyzing the waveform data of 
single-channel DAS\cite{A_new_spectral_estimation-based_feature_extraction_method_for_vehicle_classification_in_distributed_sensor_networks, 
Strategic-cum-Domestic_vehicular_movement_detection_through_Deep_Learning_approach_using_designed_Fiber-Optic_distributed_vibration_sensor}. 
While these methods are computationally efficient, they fall short in effectively incorporating 
spatial information from multiple DAS channels. Other studies have accounted for spatial data across 
different DAS channels\cite{Self-Supervised_Velocity_Field_Learning_for_High-Resolution_Traffic_Monitoring_with_Distributed_Acoustic_Sensing, 
Vehicle_Detection_and_Classification_Using_Distributed_Fiber_Optic_Acoustic_Sensing}, which are adept 
at managing scenarios with clear signals, such as multiple vehicles moving in the same direction. 
However, these methods often do not address more complex situations, like those involving two-way 
vehicle traffic with low signal-to-noise ratios. Some research\cite{Distributed_Acoustic_Sensing_for_Vehicle_Speed_and_Traffic_Flow_Estimation, 
Next-Generation_Traffic_Monitoring_with_Distributed_Acoustic_Sensing_Arrays_and_Optimum_Array_Processing} 
has ventured into these complex real-world conditions. Although these advanced methods can estimate 
vehicle count and speed amidst noise interference, they struggle to precisely capture the waveform 
trajectories produced by the vehicles.

To establish an efficient and robust vehicle monitoring system using DAS, we propose a comprehensive 
deep learning solution encompassing the creation of synthetic training datasets, the construction and 
training of convolutional neural networks (CNNs), and practical application. Our approach begins with 
the generation of a diversified training dataset, consisting of 200 two-dimensional sample pairs, which 
captures the essence of vehicular patterns and various disturbances present in DAS data. We employ 
this dataset to train a two-tiered deep learning strategy to perform instance segmentation of vehicle 
trajectories within DAS data and to accurately estimate their velocities.

The initial phase employs a CNN to implement binary image segmentation techniques, isolating linear 
image features despite significant DAS noise-comparable to edge detection. This phase pinpoints all 
vehicle trajectories, although they are often fragmented, incomplete, and have associated noise. 
However, it does not segregate individual vehicle paths. To overcome this limitation, the second 
phase converts these linear features into a Hough domain representation using the Hough transform
\cite{Method_and_means_for_recognizing_complex_patterns}, which illustrates the distribution of energy 
clusters, with each prominent cluster representing a distinct vehicle trajectory in the original 
spatiotemporal context.

Within the Hough domain, we deploy another CNN to refine and focus these clusters into discrete points 
and identify the most energetically significant ones. When retranslated into the spatiotemporal 
domain, each point denotes an individual, coherent vehicle trajectory. Additionally, the angular 
position of each point within the Hough domain is directly convertible into the vehicle's speed, 
facilitating the demarcation of individual vehicle trajectories (instance segmentation), preserving 
the integrity of these trajectories, and ascertaining the speed of each vehicle. The strategic use of 
local linearity priors for vehicle trajectories through the Hough transformation is pivotal to the 
success of the second phase. Our methodology has been rigorously tested across various real-world 
scenarios, demonstrating its effectiveness in vehicle monitoring.

\begin{figure*}[!t]
\centering
\includegraphics[width=1.0\textwidth]{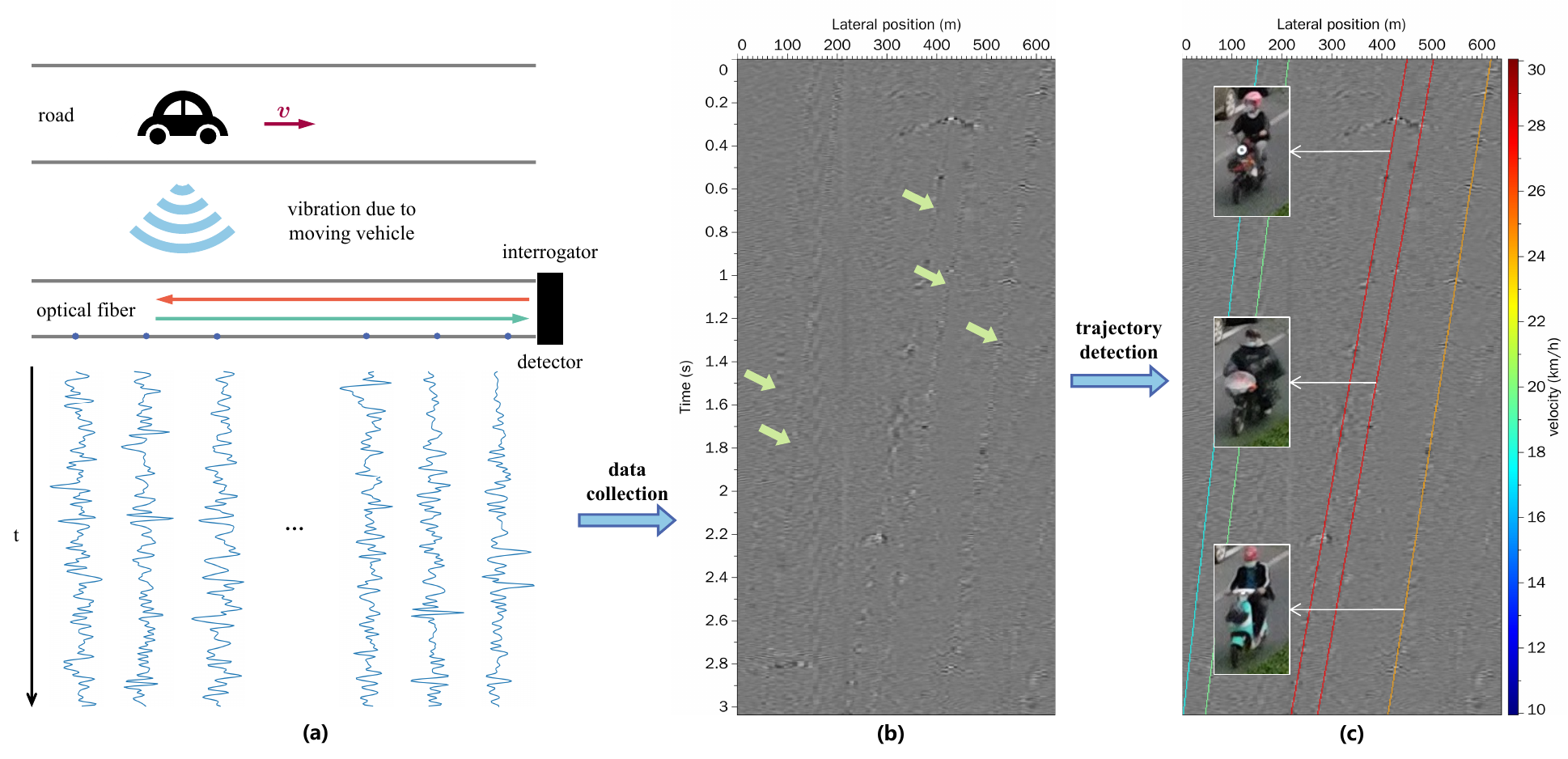}
\caption{Illustration of DAS-based traffic monitoring: (a) DAS data collection process
\cite{Distributed_Acoustic_Sensing_for_Vehicle_Speed_and_Traffic_Flow_Estimation}; (b) a collected 
real DAS data gathered in the time-space domain where the vehicle trajectories (denoted by green arrows) 
appear as linear features and are significantly blurred by noisy features; (c) individual vehicle 
trajectories (colorful lines) and their associated speeds (indicated by the color of the lines) are 
automatically estimated by our deep-learning-based method, the estimated trajectories and speeds are 
verified by the video monitoring on the same road.}
\label{cityroad_example}
\end{figure*}

\section{Problem Statement}

\mbox{Fig.~\ref{cityroad_example}} shows the entire DAS-based traffic monitoring system, which 
includes the data observation system (\mbox{Fig.~\ref{cityroad_example}a}), the data collected 
(\mbox{Fig.~\ref{cityroad_example}b}), and the instance vehicle trajectory segmentation and speed 
estimation (\mbox{Fig.~\ref{cityroad_example}c}) based on the data. The optical fiber of the DAS system 
is laid adjacent to and parallel to the roadway. One end of the fiber is connected to an interrogator, 
which sends laser pulses into the fiber and receives its backscattered light at various locations. 
As vehicles travel over the road, their induced vibrations alter the strain rate of the optical fiber, 
consequently causing a phase shift in the backscattered light. This phase shift can be received by 
the receiver to calculate the vibration of the corresponding position.

By splicing together the waveform time series obtained by the DAS system according to their positions, 
we can obtain original DAS data, as shown in \mbox{Fig.~\ref{cityroad_example}b}. The horizontal 
axis in the figure represents the position of the detection point, and the vertical axis represents 
time. Typically, the direction of travel of the vehicle is parallel to the optical fiber, and the 
speed at which the vehicle travels remains substantially constant. Therefore, the trajectory 
generated by the vehicle in the DAS data is generally some straight lines\cite{A_vehicle_detector_based_on_notched_power_for_distributed_acoustic_sensing}, 
and the speed calculated in the data space is the actual vehicle speed. Our goal is to detect the 
linear vehicle trajectories, separate them individually, and estimate their slopes (corresponding 
to vehicle velocities), as shown in \mbox{Fig.~\ref{cityroad_example}c}.

\section{Methods}

\begin{figure*}[!t]
\centering
\includegraphics[width=1.0\textwidth]{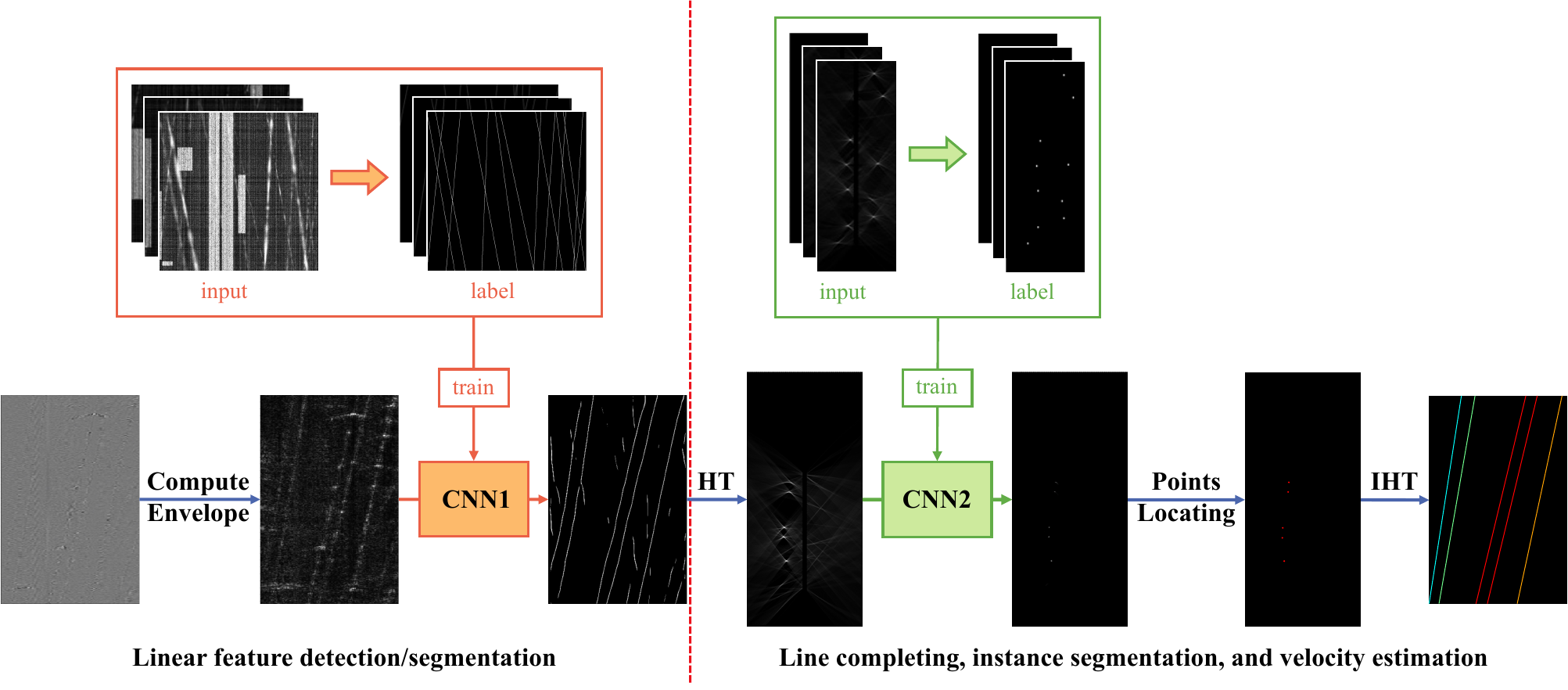}
\caption{Illustration of our two-step workflow. In the first step on the left-hand side, we 
automatically simulated numerous synthetic DAS data and their corresponding vehicle trajectory labels 
to train a CNN to automatically detect linear features in field DAS data. In the second step on the 
right-hand side, we convert the previously detected linear features into the Hough domain and train 
another CNN to focus and locate points that are further converted back to the original time-space 
domain to finally obtain complete and separate vehicle trajectories and simultaneously estimate vehicle 
speeds (denoted by the color of the lines).}
\label{workflow}
\end{figure*}

Our proposed data processing flow mainly consists of two parts: the first step (left part of 
\mbox{Fig.~\ref{workflow}}) of CNN-based linear feature segmentation in the original space domain and 
the followed post-processing (right part of \mbox{Fig.~\ref{workflow}}) of the detected linear features 
in the Hough domain to estimate clean and complete vehicle trajectories and their associated velocities.

\subsection{CNN-based linear feature segmentation in time-space domain}

\begin{figure*}[!t]
\centering
\includegraphics[width=1.0\textwidth]{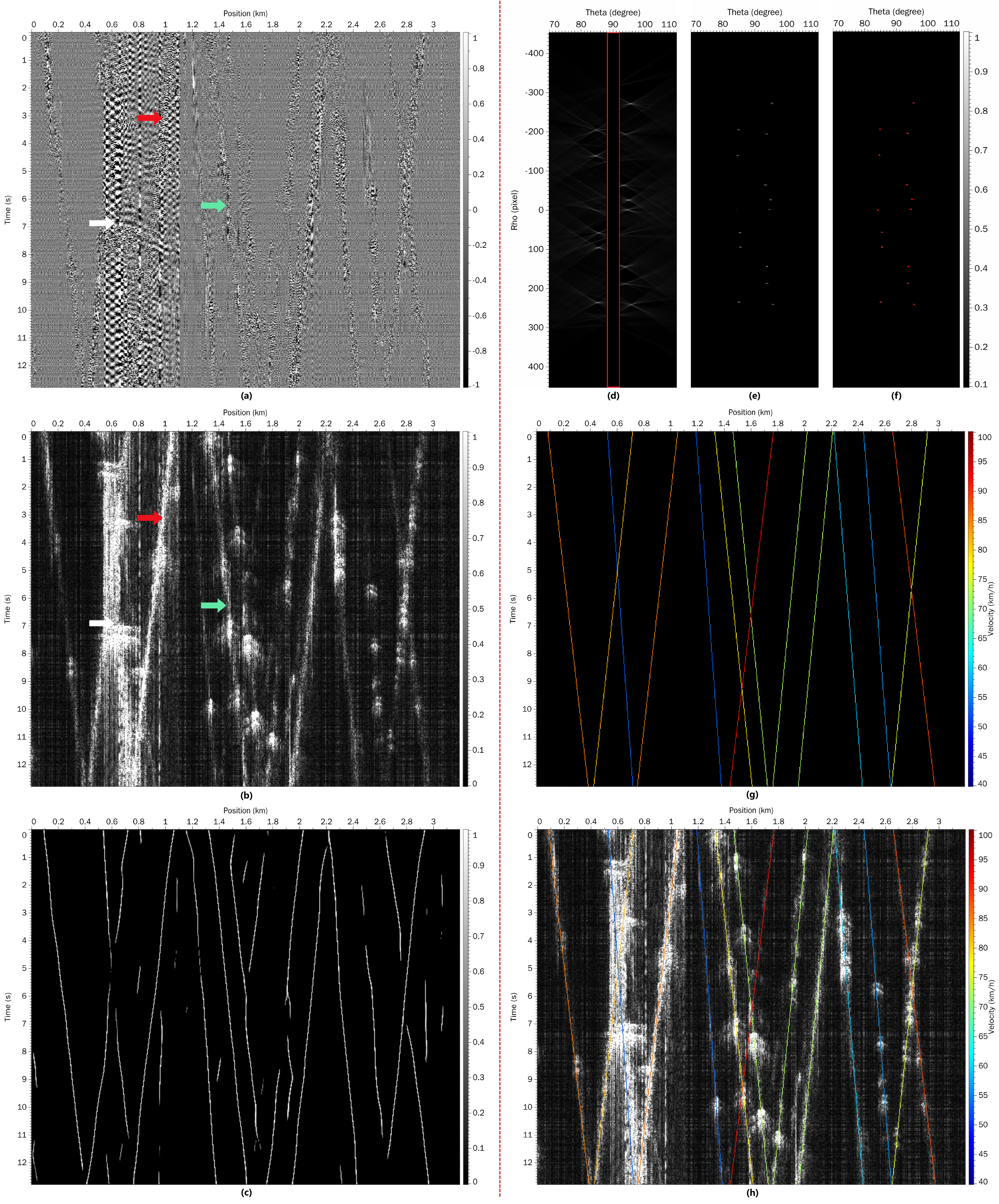}
\caption{Application of our entire intelligent workflow of traffic monitoring to a field DAS data: (a) 
original DAS data, the white arrow indicates noise, the green arrow indicates a vehicle trajectory, the 
red arrow indicates a vehicle trajectory submerged in noise; (b) an envelope image computed for (a), 
where the vehicle trajectories (denoted by the green and red arrows) are easier to resolve than in the 
original DAS data; (c) output of CNN1 for (b); (d) result of Hough transform for (c), the valued within 
the red rectangle are set to zeros to mask out static noise corresponding to nearly vertically 
distributed features in the original time-space domain; (e) output of CNN2 for (d); (f) result of 
points locating; (g) result of inverse Hough transform for (f); (h) overlay display of the finally 
picked individual vehicle trajectories and their associated speeds (denoted by the color of the lines) 
with the envelope image.}
\label{expressway_example_origin}
\end{figure*}

As shown in \mbox{Fig.~\ref{expressway_example_origin}a}, the original DAS data is highly noisy, 
where the vehicle signals or trajectories (like the one denoted by the green arrow) are significantly 
covered by strong noise features, especially those vertically distributed interferences (like the one 
denoted by the white arrow). The red arrow in the figure indicates a vehicle trajectory submerged in 
noise, we observe that the effective signal of the vehicle trajectory is weaker than the noise. 
Therefore, instead of detecting vehicle trajectories directly from raw data, we first do some 
preprocessing of the data to highlight effective vehicle trajectory features.

\subsubsection{Data Preprocessing}

First, we perform a clipping operation on the original data to remove peak noise. Subsequently, we 
compute the waveform envelope of each time series $x(t)$ in the original DAS data separately. 
The formula for calculating envelope is as follows:

\begin{equation}
\label{envelope}
E(t) = \sqrt{x^2(t)+\mathcal{H}^2[x(t)]},
\end{equation}
where $\mathcal{H}(\cdot)$ represents Hilbert transform, and $E(t)$ is the calculated envelope of $x(t)$.
\mbox{Fig.~\ref{expressway_example_origin}b} shows the envelope image calculated where the features of the 
vehicle trajectories are. Through comparison, it can be observed that the signal lines in the enveloped image 
are more obvious than the corresponding original DAS image (\mbox{Fig.~\ref{expressway_example_origin}a}), 
especially the signals in the noise indicated by the red arrow. In addition, the numerical value of a 
point in the enveloped image represents the energy of that point, making it easier to simulate compared 
to complicated waveforms in building synthetic training datasets.

\subsubsection{CNN1}
\begin{figure*}[!t]
\centering
\includegraphics[width=1.0\textwidth]{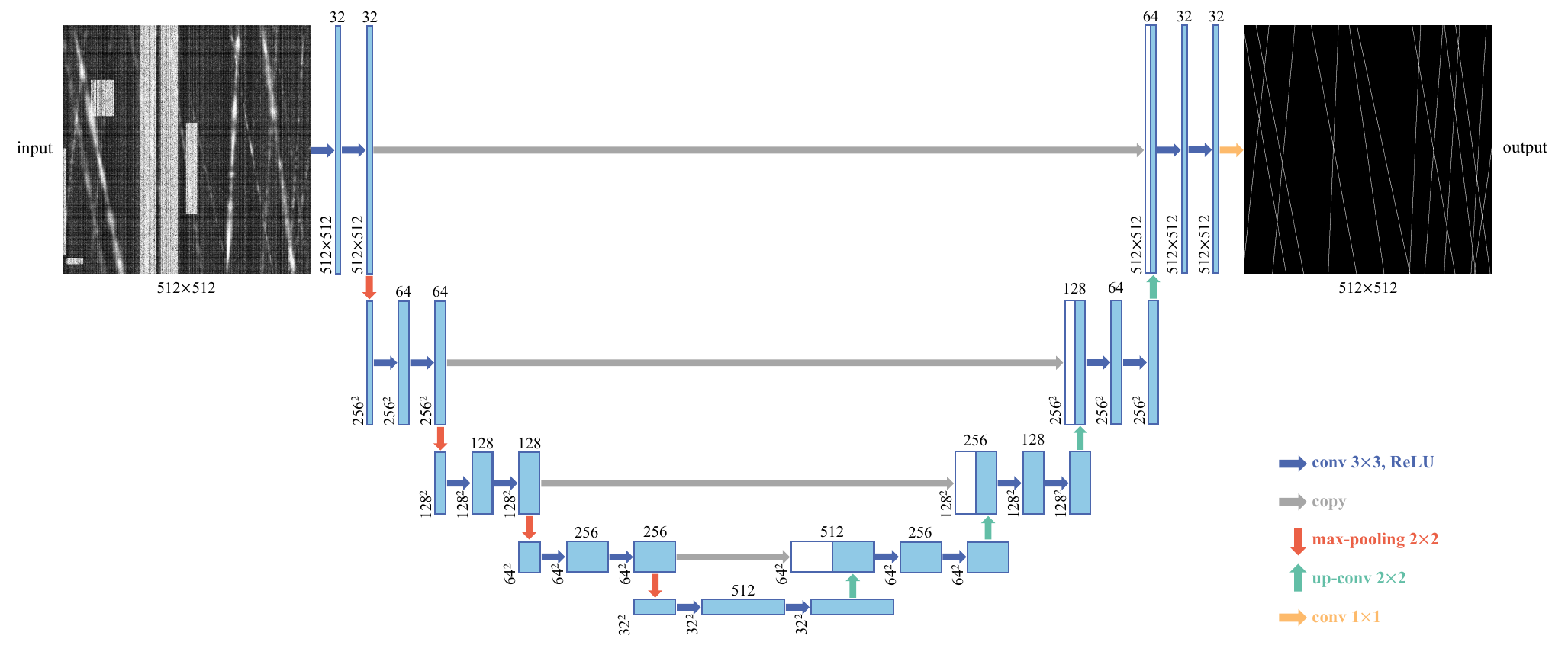}
\caption{The architecture of CNN1, which is a U-Net\cite{U-Net:_Convolutional_Networks_for_Biomedical_Image_Segmentation}. 
Each blue rectangle represents a multi-channel feature map, with the number of channels and the size of 
the feature map indicated on the top and side of the rectangle, respectively. White rectangles represent 
copied feature maps. The arrows represent various operations.}
\label{CNN1}
\end{figure*}

In the DAS envelope image within a local time window, the vehicle trajectories appear as dipping lines 
as denoted by the green arrows in \mbox{Fig.~\ref{expressway_example_origin}b} We, therefore, consider 
the task of vehicle trajectory detection as a line segmentation problem. We propose to solve such 
segmentation problem by using the widely used U-shaped CNN architecture\cite{U-Net:_Convolutional_Networks_for_Biomedical_Image_Segmentation} 
as shown in \mbox{Fig.~\ref{CNN1}}.

The structure of U-Net is described in detail below. It consists of a convolutional neural network with 
a contracting path (left side) and an expansive path (right side). The contracting path has $4$ 
downsampling modules, each comprising two repeated applications of a $3\times3$ convolution, followed 
by a rectified linear unit (ReLU), and a $2\times2$ max pooling operation with a stride of $2$ for 
downsampling. In each downsampling module, the number of feature channels is doubled. The expansive 
path also has $4$ upsampling modules, each including the upsampling of the feature map, followed by 
a $2\times2$ convolution (“up-convolution”) to halve the number of feature channels, concatenation 
with the corresponding feature map from the contracting path, and two $3\times3$ convolutions, each 
followed by a ReLU. In the last layer, a $1\times1$ convolution is used to map each $32$-dimensional 
feature vector to $1$ dimension\cite{U-Net:_Convolutional_Networks_for_Biomedical_Image_Segmentation}.

\subsubsection{Building synthetic training dataset for linear trajectory segmentation}

For the proposed CNN-based linear feature segmentation task, training data is crucial. However, in 
practice, obtaining a labeled training dataset is challenging, and manual annotation of such datasets 
is labor-intensive and subjective. To address the issue of a lack of training samples, we propose a 
comprehensive workflow to construct a large and diverse labeled sample library.

\begin{figure*}[!t]
\centering
\includegraphics[width=1.0\textwidth]{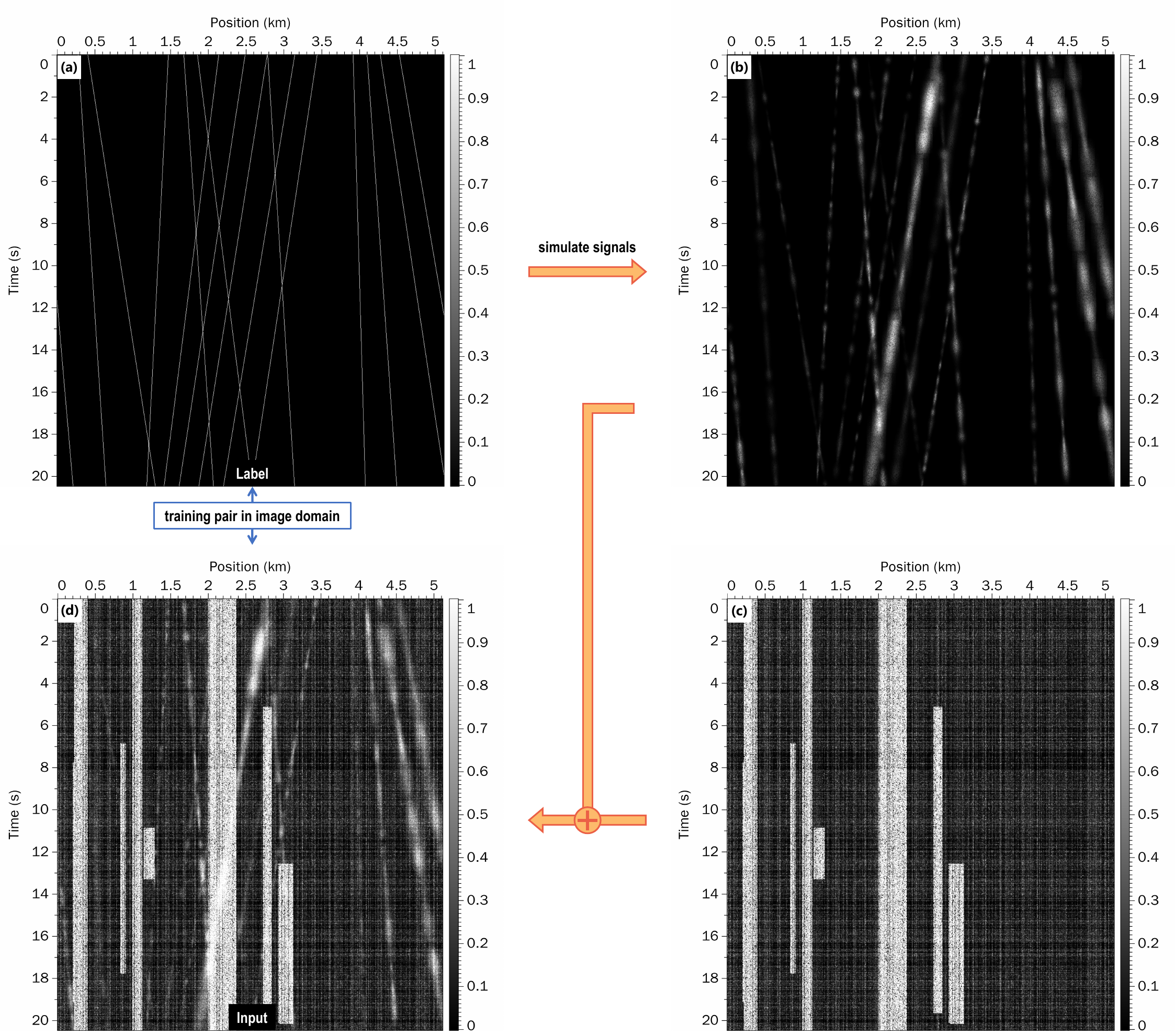}
\caption{Our workflow of simulating synthetic DAS data and their corresponding vehicle trajectory labels 
for training the CNN1 of linear trajectory segmentation in field DAS data. We first randomly generate 
some vehicle trajectories (a) as training labels, then use a series of anisotropic Gaussian kernels to 
simulate spatially and temporally varying vehicle vibrations to obtain realistic vehicle trajectory 
features in (b), we also generate noisy features in (c) and combine (b) and (c) to obtain a finally 
simulated envelope image of DAS data (d). (d) and (a) forms a training pair of input and label, 
respectively, for the CNN1.}
\label{dataset_image}
\end{figure*}

In this process, we start generating a training label as follows: we create a $512\times512$ matrix 
filled with zeros, then randomly choose a point and assign a slope within a certain range. This creates 
a straight line representing the trajectory of a vehicle in the time-space domain. The sign of the 
slope indicates the direction of the vehicle's movement, and the magnitude represents the speed. 
The range of the slope is determined based on the actual speed range in the scenario. We randomly 
choose the number of lines, denoted as n, within a certain range. Using this method, we randomly 
generate n straight lines in the matrix, setting the values at the positions where the lines pass 
through to 1, thus obtaining a label, as shown in \mbox{Fig.~\ref{dataset_image}a}.

Then, based on the generated label, we synthesize the input training data, mainly consisting of 
simulating vehicle signals and noise. To mimic the varying vibration intensities of vehicles in real 
scenarios, we assign a random average energy of less than 1 to each path in the label. As the vibration 
strength at different positions along the same path varies with the coupling between the vehicle, the 
ground, and the physical properties of the medium, we use anisotropic Gaussians arranged along the path 
to simulate the non-stationary decay of signal energy. The average energy of each Gaussian kernel 
fluctuates around the average energy of the path, and the decay coefficient and the distance between 
the centers of the two Gaussian kernels are randomly assigned within certain ranges. Additionally, we 
introduce a random perturbation to each point, i.e., multiplying the result of the previous simulation 
by a coefficient from a Gaussian distribution with a value less than 1, to better simulate 
non-stationarity. The simulated results of vehicle signals obtained by the above method are shown in 
\mbox{Fig.~\ref{dataset_image}b}. Then we simulate the noise. There are mainly two types of noise in 
actual data, one being noise blocks with strong energy, and the other being fine noise lines. Noise 
blocks typically distribute along the vertical direction and have strong energy, usually caused by 
persistent noise in a certain region or energy accumulation due to poor coupling between instruments 
and the underground medium. Noise lines distribute along both vertical and horizontal directions, 
have weaker energy, and are densely arranged, usually generated by vibrations at the location of the 
demodulator. To simulate noise blocks, we randomly determine the number, position, width, and average 
energy of noise blocks, and then similarly introduce a Gaussian perturbation to each point within the 
noise block. To simulate noise lines, we randomly generate some vertical and horizontal lines 
throughout the matrix, where the number, position, and average energy of the lines are random, and 
each line has a random perturbation as well. The average energy of noise blocks is generally greater 
than that of noise lines. The simulated results of noise obtained by the above method are shown in 
\mbox{Fig.~\ref{dataset_image}c}. Finally, we add the simulated results of vehicle signals and noise to 
obtain an input training data, as shown in \mbox{Fig.~\ref{dataset_image}d}.

\subsubsection{Training and testing}

We generated 200 training sample pairs using the aforementioned method, using 80\% as the training 
set and the remaining 20\% as the validation set. To ensure a similar distribution of input data for 
improved network generalization, we applied min-max normalization to the data (including training, 
validation, and real data) before feeding it into the network. We chose Mean Squared Error (MSE) as 
the loss function, utilized the Adam optimizer to minimize the loss function, set epoch=100, batch 
size=8, and used a Tesla A100 GPU to accelerate network training, resulting in a well-trained model 
named CNN1.

\begin{figure*}[!t]
\centering
\includegraphics[width=1.0\textwidth]{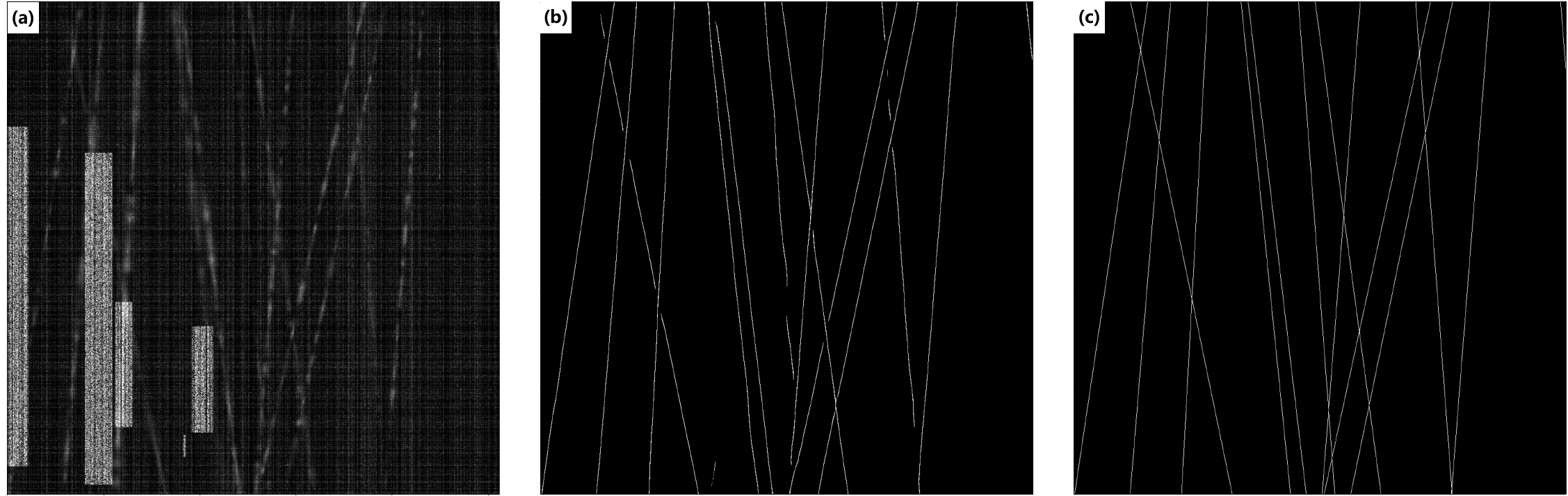}
\caption{The performance of CNN1 on a validation data. (a) is the input of validation data, (b) is the 
output after passing through CNN1, and (c) is the corresponding label.}
\label{validate_image}
\end{figure*}

We validate the performance of the trained model using validation data: \mbox{Fig.~\ref{validate_image}} 
respectively show a validation input data, its output after passing through CNN1, and its corresponding 
label. We can observe that CNN1 can generally identify all vehicle trajectories in this validation input 
data. We also test the model on the field DAS data (\mbox{Fig.~\ref{expressway_example_origin}b}, the 
envelope map computed from the original DAS data in \mbox{Fig.~\ref{expressway_example_origin}a}) and 
obtain the detection result shown in \mbox{Fig.~\ref{expressway_example_origin}c}. We can observe that 
the network accurately extracts most vehicle signals while ignoring a significant portion of the noise. 
However, there is room for improvement because some vehicle trajectories are disjointed and incomplete 
(especially in areas where multiple trajectories intersect) and some noisy linear features (unreal 
trajectories) are also detected. Additionally, different vehicle signals are not effectively separated, 
making it challenging to monitor individual vehicles and estimate vehicle speed based on this output.

\subsection{Line completion, instance segmentation and velocity estimation in Hough domain}

\subsubsection{Hough Transform}

For the reasons mentioned above, we consider further processing the output of CNN1 to achieve line 
completion, instance segmentation, and velocity estimation. Within a small time window, we can 
assume a vehicle moves straightly with an approximately uniform speed, therefore we can assume a 
vehicle trajectory in a small time window is an approximately straight line. With this assumption, 
we consider introducing the line constraint into the line completion, instance segmentation and 
velocity estimation by using the Hough transform. Its general idea is to transform the previously 
detected linear features (\mbox{Fig.~\ref{expressway_example_origin}c}) from the time-space domain 
to the Hough domain, perform energy focusing and point locating in the Hough domain, and finally 
transform the located individual points of the Hough domain back to the original time-space domain 
to obtain complete and separate lines.

\begin{figure}[!t]
\centering\includegraphics[width=0.48\textwidth]{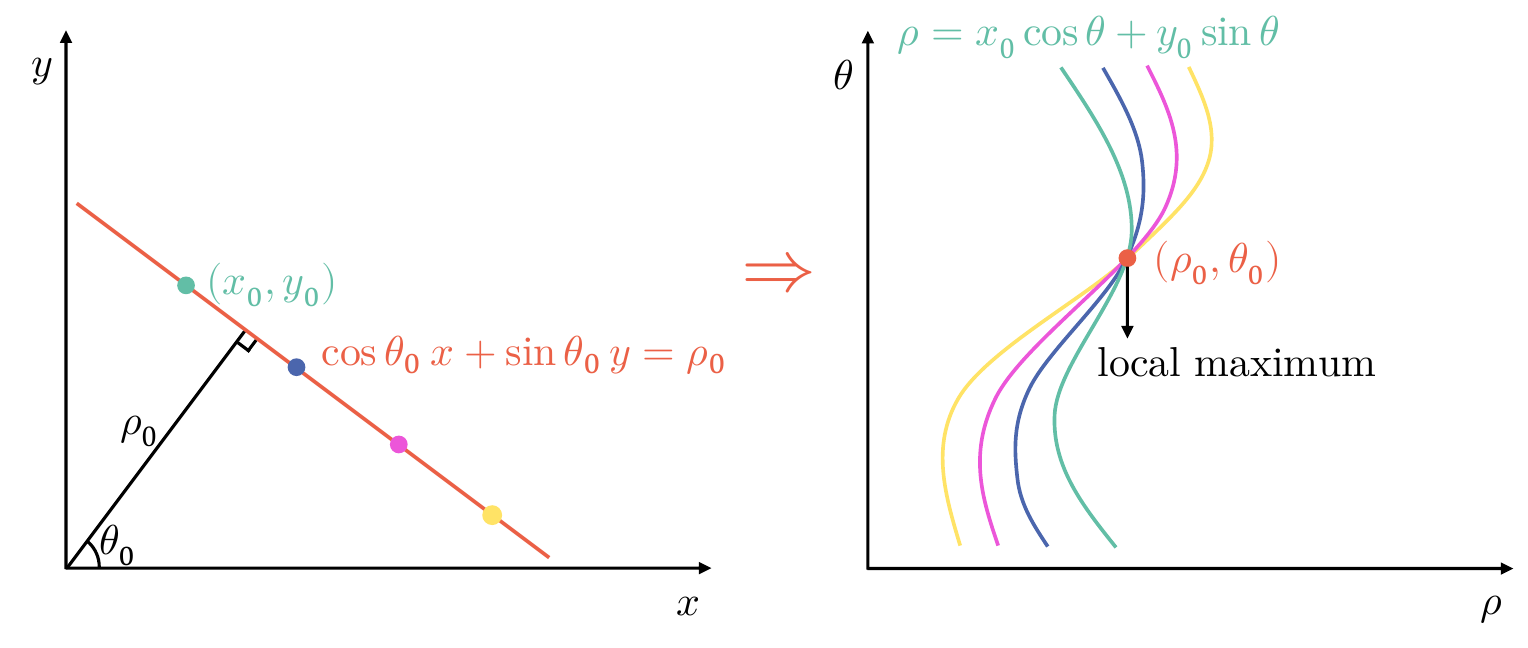}
\caption{Schematic diagram of Hough transform.}
\label{HT}
\end{figure}

\begin{figure}[!t]
\centering\includegraphics[width=0.48\textwidth]{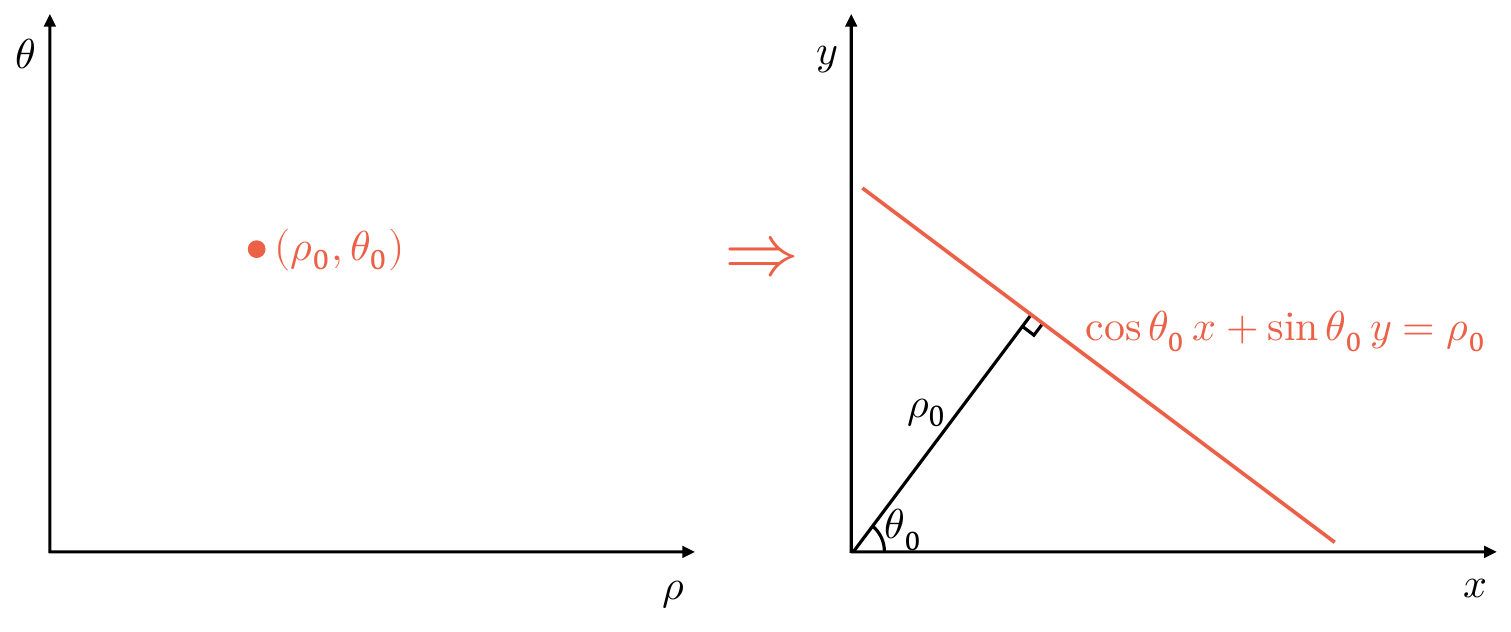}
\caption{Schematic diagram of inverse Hough transform.}
\label{IHT}
\end{figure}

The principle of the Hough transform is shown in \mbox{Fig.~\ref{HT}}. Consider a point $(x_0, y_0)$ on 
line \mbox{$l: \cos\theta_0x+\sin\theta_0y=\rho_0$} in the original data domain ($x-y$ domain), and 
transform it into a sinusoid \mbox{$\rho=x_0\cos\theta+y_0\sin\theta$} in the Hough domain 
($\rho-\theta$ domain). It is easy to prove that the point $P(\rho_0, \theta_0)$ in the Hough domain 
lies on the sinusoid \mbox{$\rho=x_0\cos\theta+y_0\sin\theta$}. We perform this transformation on every 
point on line $l$ in the original data domain and get a local maximum value point $P$ in the Hough 
domain, which indicates the parameters of line $l$ in the original data domain. In this way, we 
transform the task of detecting lines in the original data domain into detecting local maximum points 
in the Hough domain. The principle of inverse Hough transform is shown in \mbox{Fig.~\ref{IHT}}. For 
a certain point $P(\rho_0,\theta_0)$ in the Hough domain, transform it into a straight line 
\mbox{$l: \cos\theta_0x+\sin\theta_0y=\rho_0$} in the original data domain.

By applying the Hough transform to the previous linear feature detection (\mbox{Fig.~\ref
{expressway_example_origin}c}), we obtain the corresponding map in the Hough domain shown in \mbox
{Fig.~\ref{expressway_example_origin}d}. In this domain, the features appear as point-like energy 
clusters. Due to the noisy features in the previous line detection by CNN1, the clusters in the 
Hough domain are not well focused and one trajectory in the time-space domain may correspond to 
multiple maximum locations in the Hough domain, making it challenging to locate the true maximum 
points corresponding to real vehicle trajectories. We, therefore, propose to use another CNN in the 
Hough domain to focus the cluster energy for the maximum point location. Before that, we have masked 
out the obviously unreasonable features in the area (denoted by the red rectangle in \mbox{Fig.~\ref
{expressway_example_origin}d}) near $90$ degrees because these features come from static noise 
corresponding to the vertically distributed noisy features in the time-space domain (Fig. \mbox
{Fig.~\ref{expressway_example_origin}b}). This convenience of filtering out the static noise is 
another advantage of the post-processing in the  Hough domain.

\subsubsection{CNN-based point focusing and locating in the Hough domain}

To address the issue of local maxima points in the Hough domain not aligning well with target lines, 
we consider training a deep learning model CNN2 that can concentrate the originally scattered energy 
in the Hough domain onto a single point. The architecture of CNN2 is identical to CNN1, with the only 
difference being the input size and its training dataset.

\begin{figure*}[!t]
\centering
\includegraphics[width=1.0\textwidth]{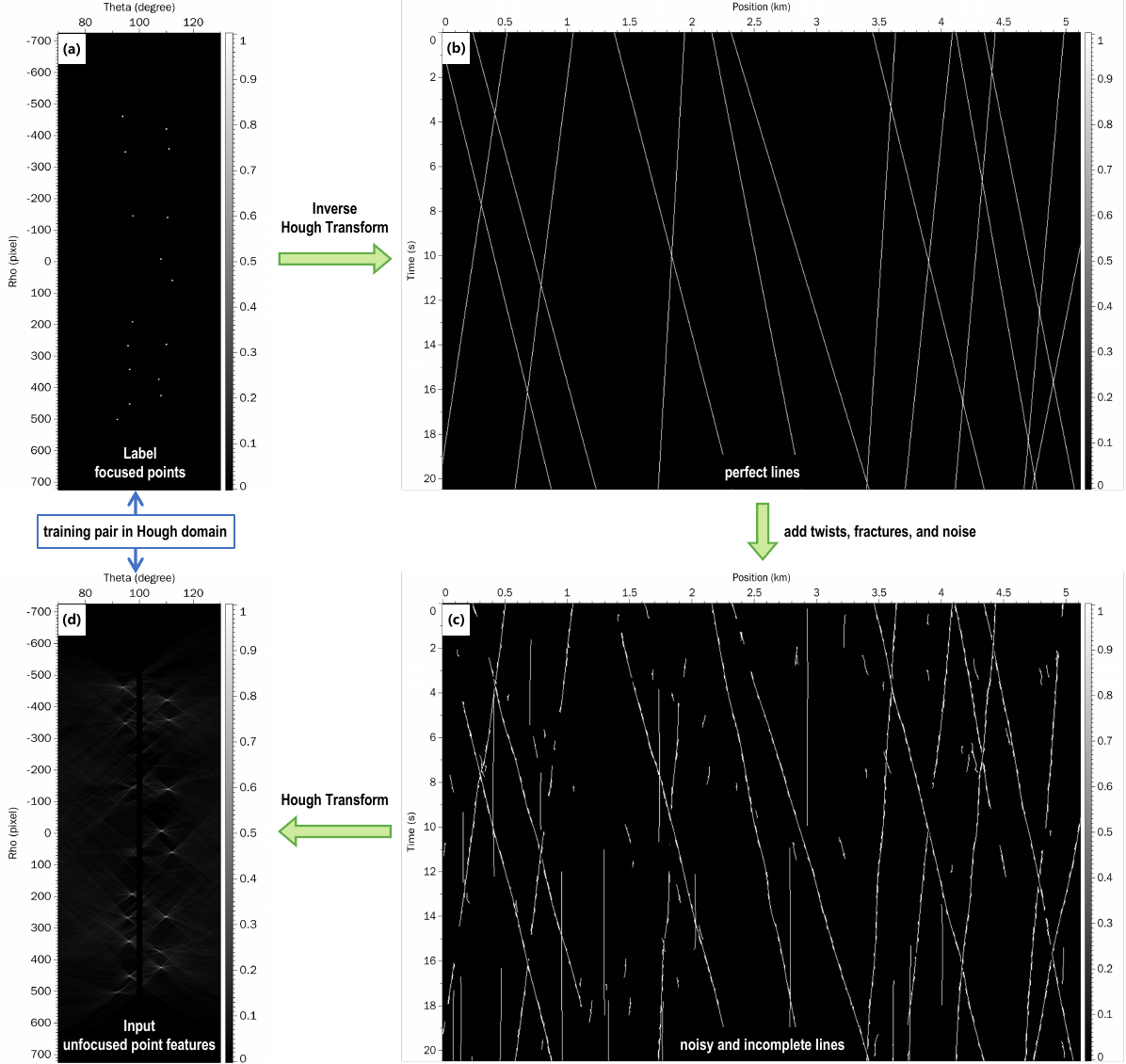}
\caption{Our workflow of simulating synthetic training dataset in the Hough domain for the CNN2: We 
first randomly generate some points in the Hough domain as training labels (a), then apply inverse 
Hough transform to (a) and obtain the corresponding vehicle trajectories in the time-space domain (b). 
After that, add some random twists, fractures, and noise to (b) to simulate the outputs of CNN1 as 
shown in (c). Finally, we apply Hough transform to (c) and obtain an image with noisy and unfocused 
point-like cluster energy features as shown in (d). (d) and (a) forms a training data pair of input and 
label, respectively, in the Hough domain to train CNN2 to focus and locate points as in the label.}
\label{dataset_htiht}
\end{figure*}

We still use synthetic data to train the network. The dataset is generated as follows: first, we 
randomly generate some points in the Hough domain as training labels (as shown in \mbox{Fig.~\ref
{dataset_htiht}a}). Secondly, we perform the inverse Hough transform on these points to obtain the 
corresponding lines in the image domain (as shown in \mbox{Fig.~\ref{dataset_htiht}b}). Then, to 
simulate the inaccuracy of the preliminary picking results (as shown in \mbox
{Fig.~\ref{expressway_example_origin}c}), we add some random curvature, discontinuity, and noise to 
these lines (as shown in \mbox{Fig.~\ref{dataset_htiht}c}). Finally, we perform the Hough transform 
on them to obtain some unfocused energy clusters, which serve as the corresponding input training 
data (as shown in \mbox{Fig.~\ref{dataset_htiht}d}).

We use the same training strategy as CNN1 to train the model CNN2. The result of the Hough transform 
(shown in \mbox{Fig.~\ref{expressway_example_origin}d}) is then input into CNN2, producing the 
result shown in \mbox{Fig.~\ref{expressway_example_origin}e}. It can be observed that compared to 
\mbox{Fig.~\ref{expressway_example_origin}d}, the energy in \mbox{Fig.~\ref
{expressway_example_origin}e} is more concentrated, achieving the picking of local maximum points.

Using the above method, we obtained a Hough domain image with relatively concentrated energy, 
characterized by many distinct clusters of points (as shown in \mbox{Fig.~\ref
{expressway_example_origin}e}). We aim to obtain separate points, so we perform localization on 
these points, aggregating groups of points that are close in distance into a single point. As shown 
in \mbox{Fig.~\ref{expressway_example_origin}f}, the red points represent the result of point 
locating, and their coordinates correspond to the detected parameters ($\rho$ and $\theta$) of the 
lines and the parameter $\theta$ can be easily converted to the slope of the line of the vehicle speed.

\begin{figure*}[!t]
\centering\includegraphics[width=1.0\textwidth]{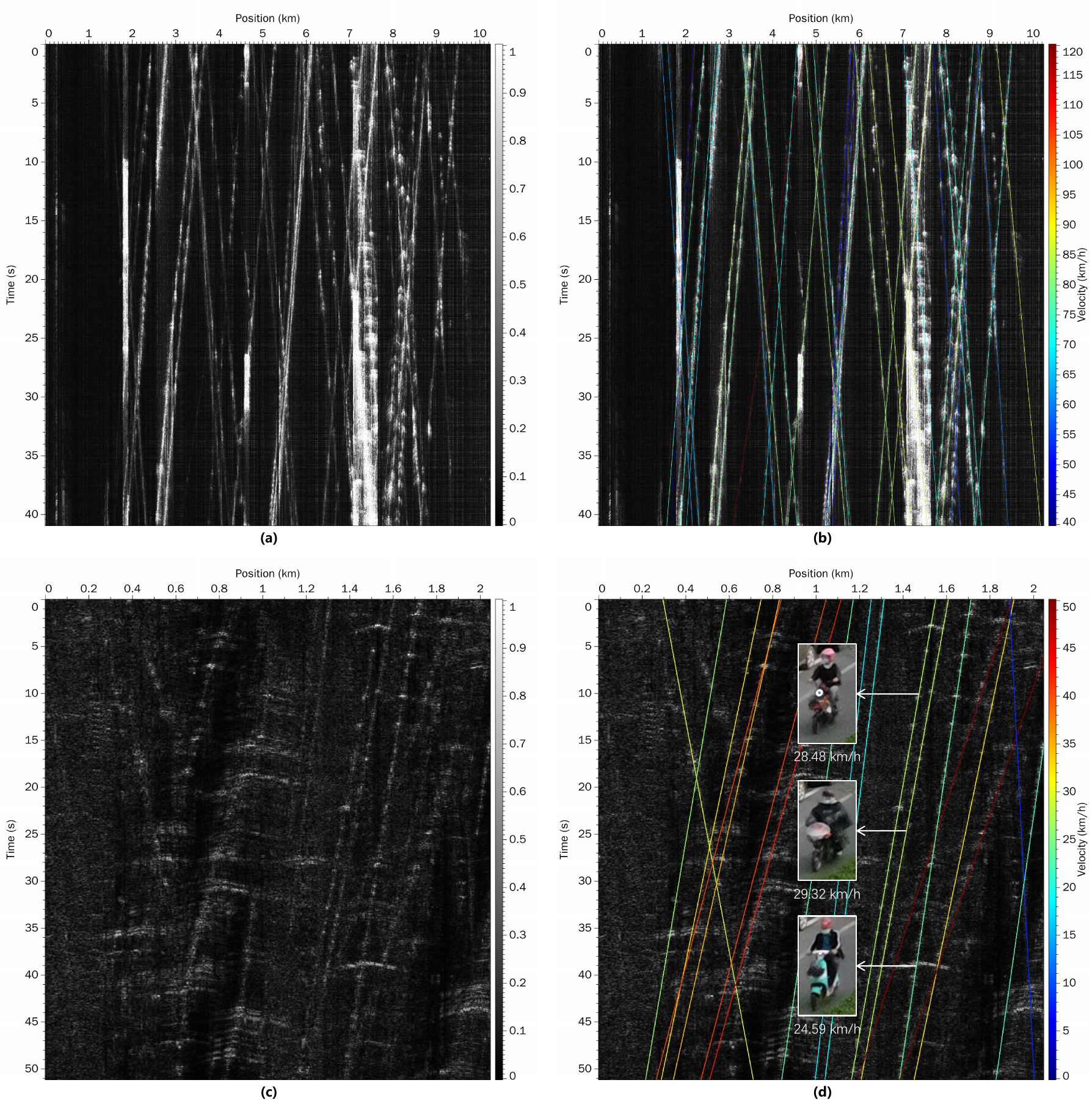}
\caption{Applying our entire intelligent traffic monitoring workflow to expressway data (a) and city road 
data (c), the obtained results are respectively shown in (b), (d).}
\label{application_result}
\end{figure*}

\subsubsection{Inverse Hough Transform}

We applied the inverse Hough transform to the result of point locating (as shown in \mbox{Fig.~\ref
{expressway_example_origin}f}), and the obtained result is presented in \mbox{Fig.~\ref
{expressway_example_origin}g}. Comparing this result with the output of CNN1 (as shown in \mbox
{Fig.~\ref{expressway_example_origin}c}), it can be observed that the vertical noise lines have 
been removed, and all the linear trajectories are complete and each of them is separated (achieving 
instance segmentation by the separate point located in the Hough domain). In addition, we also 
automatically estimate the corresponding vehicle speeds as displayed in color in \mbox{Fig.~\ref
{expressway_example_origin}g}. Overlaying this result with the envelope image (as shown in \mbox
{Fig.~\ref{expressway_example_origin}b}), the picking result is displayed in \mbox{Fig.~\ref
{expressway_example_origin}h}, where the detected lines align well with the original vehicle signals 
recorded in the DAS data.

\section{Application}

In order to test the effectiveness of the entire workflow of intelligent traffic monitoring, we apply 
it to the field data from two scenarios of expressway and city road as shown in 
\mbox{Fig.~\ref{application_result}}.

\subsection{Expressway}

Applying the above method to the data collected on an expressway (shown in the \mbox{Fig.~\ref
{application_result}a}), the results are shown in \mbox{Fig.~\ref{application_result}b}. 
It can be observed that most of the vehicle signals are identified accurately, and the vertical noise 
signal is completely removed, and the calculated speed is also within a reasonable range, which 
shows that the model we trained is effective in the expressway scene.

\subsection{City Road}

In order to further test the generalization of the model and verify the accuracy of the picking 
results, we collecte DAS data on a city road (shown in the \mbox{Fig.~\ref{application_result}c}) 
and shoot a video to record the driving information of the vehicle. We apply the above method to the 
data and the results are shown in \mbox{Fig.~\ref{application_result}d}. The photos and marked 
speeds in the figure are calculated based on the video. It can be observed that during the time period 
shown in the figure, the picking results are in good agreement with the vehicle position and speed 
recorded in the video, which indicates that the model has certain stability and cross-scenario 
generalization. 

However, there are also some errors and omissions in the result. This may be due to the fact that the 
traffic density and speed of city roads are quite different from those of expressways, the equipment 
used in city roads is different from that in expressways, and poor coupling between fiber and ground 
leads to poor data quality.

\section{Conclusions}

In this paper, we propose a DAS-based intelligent traffic monitoring algorithm that combines simulating 
synthetic DAS training datasets, deep learning for linear feature segmentation and point feature 
focusing, and Hough transform with line constraints. The simulation generates 200 synthetic training 
data pairs, which solves the common challenge of missing labeled DAS data in the scenario of traffic 
monitoring. We have made the datasets publicly available through https://github.com/TTMuTian/itm/.
Deep learning method can quickly perform initial local feature pickup on vehicle-generated DAS signals, 
and then the Hough transform helps extract global information in a parameterized form for global semantic 
extraction. Although we train the neural network with synthesized datasets, our algorithm can quickly 
and accurately identify the vast majority of vehicle-generated DAS signals on complex real-world 
scenarios as well. This algorithm is helpful in realizing low-cost, high temporal and spatial resolution 
real-time vehicle detection, and has reference significance for some similar traffic monitoring 
scenarios. In the future, we plan to conduct in-depth research on the forward modeling process of data 
and the migration of models in different scenarios, and further optimize the model structure.

The method proposed in this paper still has some limitations. The process of our data set construction 
lacks the support of more accurate physical processes. By considering more accurate physical processes 
in the forward modeling for simulating a more realistic synthetic training dataset, the deep learning 
model might be better trained to achieve better performance. The proposed two-step workflow is also 
relatively complex, which affects the detection efficiency. We may consider including the Hough 
transform as a module in the deep neural network. In this way, we might be able to combine our workflow 
of two steps into one to simplify the entire processing and improve efficiency.

\section*{Acknowledgments}
This research is financially supported by the National Key R\&D Program of China (2021YFA0716903) and the 
Fund Project of Science and Technology on Near-Surface Detection Laboratory under the grand no. 6142414211101. 
We thank the USTC Supercomputing Center for providing computational resources for this work.

% {\appendix[Proof of the Zonklar Equations]
% Use $\backslash${\tt{appendix}} if you have a single appendix:
% Do not use $\backslash${\tt{section}} anymore after $\backslash${\tt{appendix}}, only $\backslash${\tt{section*}}.
% If you have multiple appendixes use $\backslash${\tt{appendices}} then use $\backslash${\tt{section}} to start each appendix.
% You must declare a $\backslash${\tt{section}} before using any $\backslash${\tt{subsection}} or using $\backslash${\tt{label}} ($\backslash${\tt{appendices}} by itself
%  starts a section numbered zero.)}


\bibliographystyle{IEEEtran}
\bibliography{Intelligent_Traffic_Monitoring_with_Distributed_Acoustic_Sensing}

\vfill

\end{document}